\documentclass[12pt,cite,epsfig,wrapfig,subfigure]{article} 
\usepackage{epsfig}
\usepackage{cite} 
\newcommand{\eq}{\begin{equation}} 
\newcommand{\eqx}{\end{equation}} 
\newcommand{\eqn}{\begin{eqnarray}} 
\newcommand{\eqnx}{\end{eqnarray}} 
 
\newcommand{\fsn}{$g_1(x,Q^2)$} 
\newcommand{\dln}{ln$^2(1/x)$}

\newcommand{\nin}{\noindent} 
\def\lapproxeq{\lower .7ex\hbox{$\;\stackrel{\textstyle
<}{\sim}\;$}}
\def\gapproxeq{\lower .7ex\hbox{$\;\stackrel{\textstyle
>}{\sim}\;$}}
\begin{document} 
\nin 


%
%
\mbox{}
\\

\vskip2cm
\begin{center} 
{\Large \bf Spin dependent structure function $g_1(x,Q^2)$  
 at low $x$ and low~$Q^2$
}\\ 
\vspace{10mm} 
 
{\bf Barbara~Bade\l{}ek} 
\footnote {\noindent On leave of absence from the
                     Institute of Experimental Physics, Warsaw University,
                       Ho\.za 69, \\ PL-00 681 Warsaw, Poland; 
e-mail address: badelek@cern.ch
}\\ 
{European Organization for Nuclear Research, CERN, Geneva, Switzerland} \\

\vspace{1cm}
{\it Dedicated to Jan Kwieci\'nski in honour of his 65th birthday} 

\end{center}

\vspace{5mm} 
\nin 
{\bf Abstract:} 
{\footnotesize  
This is  a review of experimental and phenomenological 
investigations of the 
nucleon spin dependent structure function $g_1$ at low values of $x$ and  $Q^2$. 
 
} 
\vspace{6mm} 
\section{Introduction} 

Spin has for the first time manifested itself experimentally as a new and
non-classical quantity in the Stern-Gerlach experiment 
in 1921, essentially before the birth of the modern quantum mechanics
and before (what is being accepted as) the spin discovery.
The history of spin, \cite{martin}, and its predictable future, \cite{future}, 
are both very exciting. With spin research programmes presently operating 
at BNL, CERN, DESY, JLAB and SLAC and with prospects of polarised 
$e-p$ collider, EIC, and polarised $e^+e^-$ linear colliders
we are witnessing a wide attempt to understand the spin, test 
the spin sector of QCD and possibly also use it in the search for 
``new physics''.  

This paper is a review of results of the experimental and theoretical 
investigations of the nucleon spin structure 
at low values of the Bjorken scaling variable $x$. This is a region
of high parton densities, where new dynamical mechanisms may be revealed
and where the knowledge of the spin dependent nucleon structure function 
\fsn ~is required to 
evaluate the spin sum rules necessary to understand the origin of
the nucleon spin. The behaviour of $g_1$ at $x\lapproxeq$ 0.001 and 
in the scaling region, $Q^2 \gapproxeq$1 GeV$^2$,
is unknown due to the lack of colliders with polarised beams. 
Information about spin-averaged structure function $F_2(x,Q^2)$ in that 
region comes almost entirely from the experiments at HERA: the $F_2$ rises 
with decreasing $x$, in agreement with QCD and the rise
is weaker with decreasing $Q^2$, \cite{heraf2}.
However even if such an inclusive quantity as $F_2$ can be described
by the conventional DGLAP resummation, certain non-inclusive observables
seem to be better described by the BFKL approach, \cite{lowx_coll}. Thus
non-inclusive reactions are crucial to understand the dynamics of high parton 
densities.
Unfortunately in the case od spin, the longitudinal structure function, \fsn,
is presently
the only observable which permits the study of low $x$ spin dependent processes.
Since it is being obtained exclusively from fixed-target experiments where 
low values of $x$ are correlated with low values of $Q^2$, 
one faces new complications: not only the measurements  
put very high demands on event triggering and reconstruction but also theoretical
interpretations of the results require a suitable
extrapolation of parton ideas to the low $Q^2$ region and inclusion of
dynamical mechanisms, like the Vector Meson Dominance (VMD). The latter may
indeed be important apart of the partonic contributions as it is the case
for the low $Q^2$ spin-averaged electroproduction,
see e.g. \cite{breitweg,el89,el92}. In the spin-dependent case and
in the $Q^2$=0 limit $g_1$ should be a finite function of $W^2$,
free from any kinematical singularities or zeros. 
For large $Q^2$ the VMD contribution to $g_1$ vanishes as $1/Q^4$
 and can usually be neglected.
The partonic contribution to $g_1$
 which controls the structure functions in the deep inelastic domain
 and which scales there {\it modulo} logarithmic corrections, has to be
 suitably extended to the low $Q^2$ region.

\section{Results of measurements}

%
%
%
%
Experimental knowledge on the longitudinal spin dependent structure 
function $g_1(x,Q^2)$ comes entirely from the
fixed-target setups: EMC, SMC and 
COMPASS at CERN, experiments at SLAC 
(E142, E143, E154, E155, E155X)
and the HERMES experiment at HERA $ep$ collider. 
Information on the kinematic variables comes from measurements of 
the incident and scattered leptons. Hadrons resulting
from the target breakup are often also measured, and -- in the case 
of HERMES and COMPASS  -- identified, if their momenta are larger than 1 GeV
in the former- and larger than 2.5 GeV in the latter case.

\begin{figure}[ht]
\begin{center}
\epsfig{figure=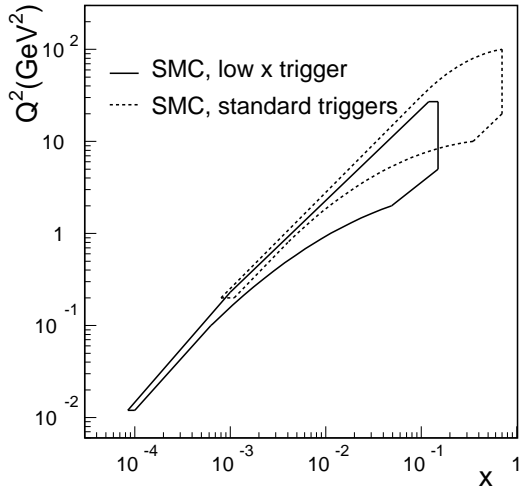,height=7cm}
\end{center}
\caption{\label{fig_t15}\footnotesize{Contours of the kinematic acceptance 
in the $(x,Q^2)$
plane for the standard triggers (dotted line) and for the low $x$ trigger 
(solid line) in the SMC.
Figure taken from \protect\cite{ssmc98}.}
}
\end{figure}

\begin{figure}[ht]
\begin{center}
\epsfig{figure=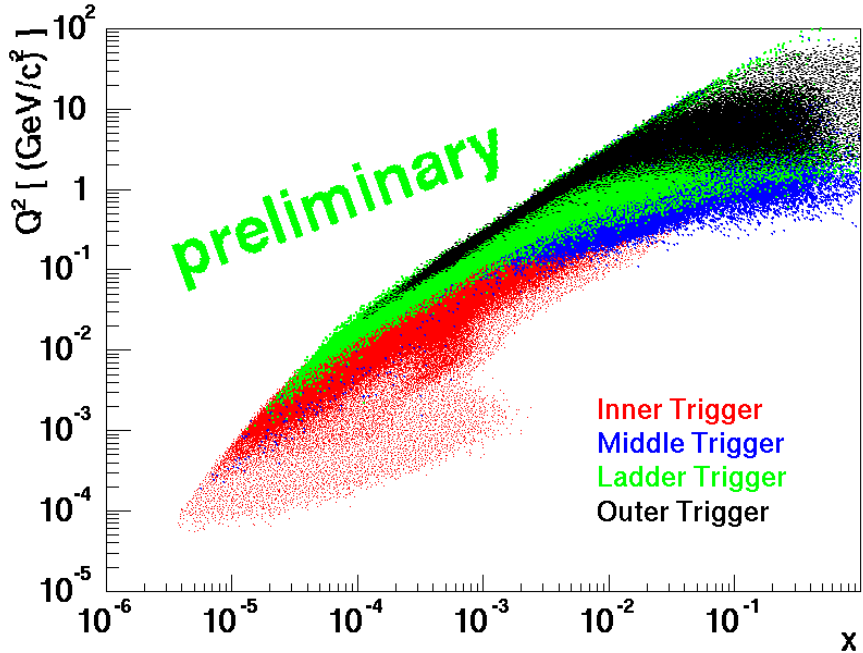,height=9cm}
\end{center}
\caption{\label{fig_compass}\footnotesize{ Contours of the kinematic 
acceptance in the $(x,Q^2)$
plane for the COMPASS triggers. Only about 5\% of data taken in 2002 are
marked. Figure taken from \protect\cite{claude}.}
}
\end{figure}

In fixed-target experiments the low $x$ region is correlated
with low values of $Q^2$ and the range of $Q^2$ covered at low $x$ is usually
limited. In the past the lowest values of $x$ were reached by the SMC 
due to a high energy of the muon beam and to a demand of
a final state hadron, imposed either in the off-line analysis \cite{smc98} or 
in the dedicated low $x$
trigger with a hadron signal in the calorimeter \cite{ssmc98}. These requirements 
permitted measurements of 
muon scattering angles as low as 1 mrad, Fig.\ref{fig_t15} and
efficiently removed the dominant  background of muons scattered 
elastically from target atomic electrons at $x=$0.000545, cf. \cite{ssmc98}.
Much lower values of $x$ are presently being obtained by COMPASS, 
Fig.\ref{fig_compass}, thanks to a specially designed trigger system, \cite{claude}.
Charged lepton deep inelastic scattering experiments benefit from high rates and
low (albeit complicated) systematic biases. They have to deal
with a strong $Q^2$ dependence of the cross section (due to photon 
propagator effects) and with large contribution of radiative processes.
Electron and muon measurements are complementary: the former offer
very high beam intensities but their kinematic
acceptance is limited to low values of $Q^2$ and moderate values
of $x$, the latter extend to higher $Q^2$ and
to lower values of $x$ (an important aspect in the study 
of sum rules) but due to limited muon beam intensities the data
taking time has to be long to ensure a satisfactory statistics.


Spin-dependent cross sections are only a small contribution to the total deep
inelastic cross section. Therefore they can best be determined by mesuring
the cross section asymmetries in which spin-independent contributions
cancel. 
Direct result of all measurements is thus the
longitudinal cross section asymmetry, $A_{\|}$ which permits to
extract the virtual photon -- proton asymmetry, $A_1$ and finally,
using $F_2$ and $R$, to get $g_1$. Asymmetry  $A_{\|}$ is small, thus a large 
statistics is necessary to make a statistically significant measurement.
Problems connected with evaluation of spin structure functions from the data
are described in detail in \cite{roland_exp}.

\begin{figure}[ht]
\begin{center}
\epsfig{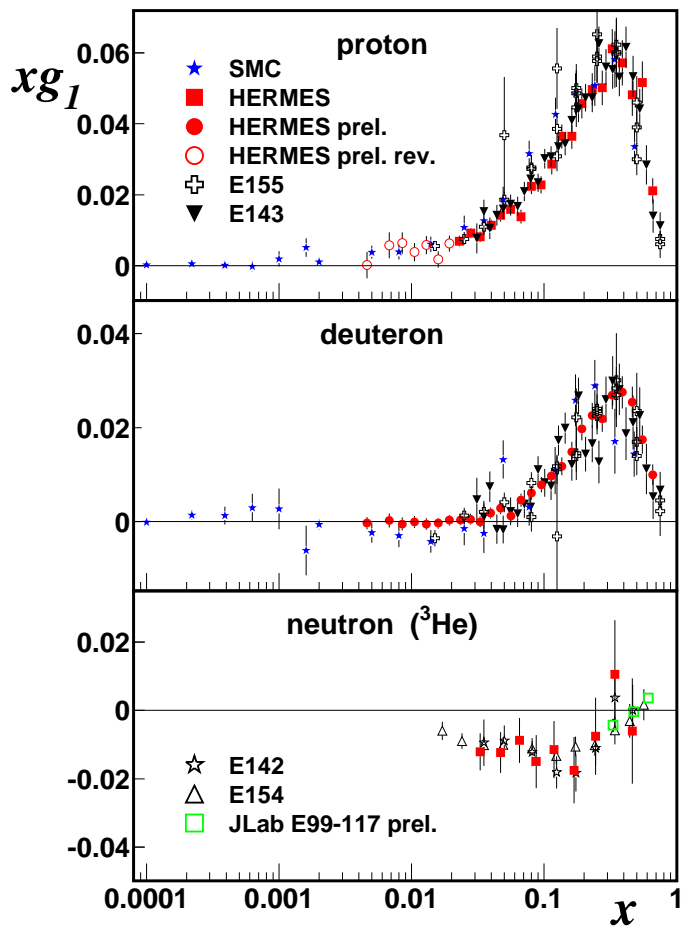}
\end{center}
\caption{\label{fig_uta}\footnotesize{Compilation of data on $x$\fsn ~including new,
preliminary data from HERMES and JLAB (E99-117). All the data
are given at their quoted mean $Q^2$ values. Errors are total.
Figure taken from \protect\cite{uta}.}
}
\end{figure}

As a result of a large experimental effort over the years, proton and
deuteron $g_1$ was measured for 0.000 06 $< x <$ 0.8, cf. Fig.
\ref{fig_uta}, \cite{uta}.
Direct measurements on the neutron are limited to $x\gapproxeq$ 0.02.
 No significant spin effects were observed at lowest values of $x$, 
explored only by the SMC. 
Scaling violation in $g_1(x,Q^2)$
is weak: the average $Q^2$ is about 10 GeV$^2$ for the SMC
and almost an order of magnitude less for the SLAC and HERMES experiments.
For the SMC data \cite{ssmc98}, $\langle x \rangle$ = 0.0001 corres\-ponds to
$\langle Q^2 \rangle$ = 0.02 GeV$^2$; $Q^2$ becomes larger than 1 GeV$^2$
at $x\gapproxeq$ 0.003 (at $x\gapproxeq$ 0.03 for HERMES).
At lowest $x$ results on $g_1$ have very large errors but it seems that
both $g_1^p$ and $g_1^d$ are positive there.
Statistical errors dominate in that kinematic interval.

\section{Regge model predictions}

The low $x$ behaviour of $g_1$ for fixed $Q^2$ reflects the high energy behaviour
of the virtual Compton scattering cross section with centre-of-mass
energy squared, $s\equiv W^2=M^2+Q^2(1/x - 1)$; here $M$ is the nucleon mass. 
This is the Regge limit of the (deep) inelastic scattering
where the Regge pole exchange model  should be applicable.
This model gives the following parametrisation of the (singlet and 
nonsinglet) spin dependent structure function at $x \rightarrow 0$ 
(i.e. $Q^2 \ll W^2$):  
\eq
g_1^i(x,Q^2) \sim \beta(Q^2) x^{-\alpha_i(0)}
\label{regge_g1}
\eqx
where the index $i$ refers to singlet ($s$) and nonsinglet ($ns$)
combinations of proton and neutron structure functions, $g_1^s(x,Q^2)=
g_1^p(x,Q^2)+g_1^n(x,Q^2)$ and $g_1^{ns}(x,Q^2)= g_1^p(x,Q^2)-g_1^n(x,Q^2)$
respectively. Intercepts of the Regge trajectories, 
$\alpha_i(0)$, are universal
quantities, independent of the external particles or currents and
dependent only on the quantum numbers of the exchanged Regge poles.
In the case of $g_1$ the intercepts correspond to the axial vector mesons
with $I$=0 ($g_1^s$; $f_1$ trajectory) and $I$=1 ($g_1^{ns}$; $a_1$ trajectory). 
It is expected that $\alpha_{s,ns}(0) \lapproxeq 0$ and that
$\alpha_s(0)\approx \alpha_ns(0)$, \cite{hei}. 
This behaviour of $g_1$ should go smoothly to the $W^{2 \alpha}$
dependence for $Q^2 \rightarrow 0$. 
A Regge type approach has been used in a global analysis of the proton and
neutron spin structure function data in the range 0.3 GeV$^2 < Q^2 <$ 70 GeV$^2$
and 4 GeV$^2 < W^2 <$ 300 GeV$^2$, \cite{nicola}; fits gave a smooth
extrapolation of $g_1$ down to the photoproduction limit.

At large $Q^2$ it is well known that
the Regge behaviour of $g_1(x,Q^2)$
is unstable against the DGLAP evolution and against 
resummation of the ${\rm ln^2}(1/x)$ terms
which generate more singular $x$ dependence than that implied by
Eq.(\ref{regge_g1}) for $\alpha_{s,ns}(0) \lapproxeq 0$, cf. Section 4. 


Other considerations based on the Regge theory give further
isosinglet contributions to the low $x$ behaviour of $g_1$: a term proportional
to {ln}$x$ (from a vector component of the short range exchange potential),
\cite{clo_rob} and a term proportional to 2~{ln}(1/$x$)--1 (exchange of two
nonperturbative gluons), \cite{bass_land}; a perversly behaving term  
proportional to 1/$(x$ln$^2x)$, recalled in \cite{clo_rob} is not valid
for $g_1$, \cite{misha}.

Testing the Regge behaviour of $g_1$ through its $x$ dependence
should in principle be possible with the
low $x$ data of the SMC \cite{ssmc98} which include the kinematic region where
$W^2$ is high, $W^2$\gapproxeq 100 GeV$^2$, and $W^2\gg Q^2$.
Thus the Regge model should be applicable there.
However for those data $W^2$ changes very little:
from about 100 GeV$^2$ at $x=$ 0.1 to about 220 GeV$^2$ at $x=$ 0.0001,
contrary to a strong change of $Q^2$: from about 20 GeV$^2$
to about 0.01 GeV$^2$ respectively.  Thus those data cannot test the Regge
behaviour of $g_1$. %
\begin{figure}[htb]
\begin{center}
\epsfig{figure=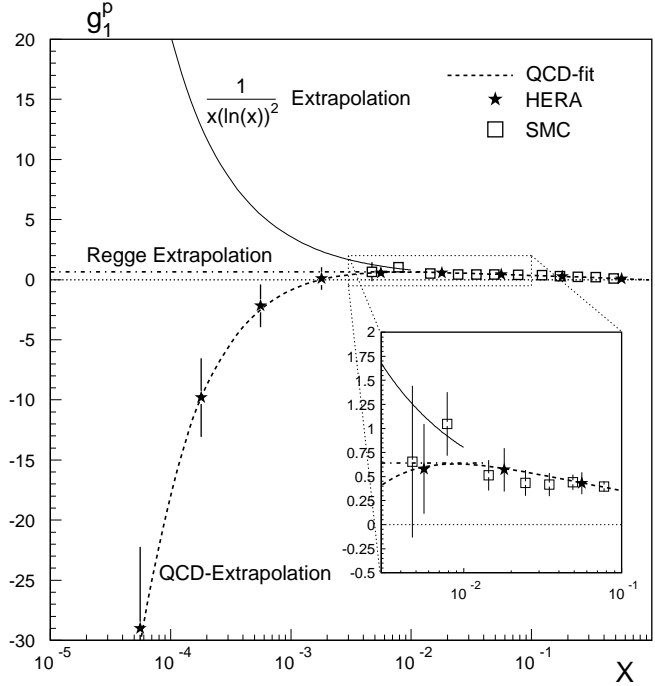,height=10cm}
\end{center}
\caption{\label{fig_g1extrapo} \footnotesize{Three scenarios of the possible 
behaviour of $g_1^p$ at low $x$ \cite{hera_study}.}}
\end{figure}
Moreover
employing the Regge model
prediction, $g_1\sim x^0$ to obtain the $x\rightarrow 0$ extrapolation of $g_1$,
often used in the past to extract the $g_1$ moments
(cf.\cite{qcd_old} and Fig.\ref{fig_g1extrapo}) is not correct. 
The values of $g_1$ should be evolved to a common value of $Q^2$ before 
the extrapolation, cf. Eq.(\ref{regge_g1}). Therefore other ways of extrapolation 
of $g_1$ to low values of $x$ were adopted in the analyses, see Sections 4.1 
and 4.3. Testing the Regge behaviour of $g_1$ may be possible in COMPASS,
cf. Fig. \ref{fig_compass}.

\section{Low $x$ implications from the perturbative QCD }   
\subsection{DGLAP fits to the $g_1$ measurements}

\begin{figure}[ht]
\begin{center}
\epsfig{figure=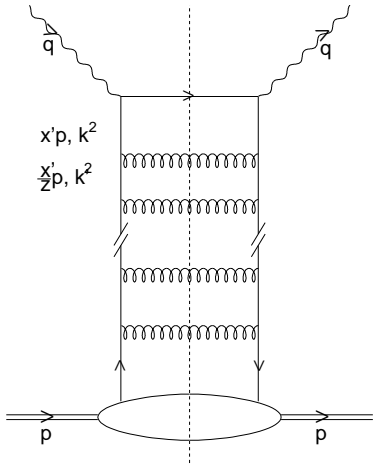,height=8cm}
\end{center}
\vspace*{-1.5cm}
\caption{\label{fig_ladder}\footnotesize{An example of a ``ladder'' diagram.
Figure taken from \protect\cite{bbjk}.}
}
\end{figure}

In the standard QCD, the asymptotic, small $x$ behaviour of $g_1$ is created by the
``ladder'' processes, Fig.\ref{fig_ladder}. In the LO approximation 
it is given by:

\eq
g_1(x,Q^2)\sim {\rm exp}\left[A\sqrt{\xi(Q^2){\rm ln}(1/x)}\right]
\label{asympt_g1}
\eqx
where
\eq
\xi(Q^2)=\int_{Q^2_0}^{Q^2}{dq^2\over q^2}{\alpha_s(q^2)\over 2\pi}
\eqx
and the constant $A$ is different for singlet and non-singlet case. 
The above behaviour of $g_1$ is more singular than that implied by
Eq.(\ref{regge_g1}) for $\alpha_{s,ns}(0) \lapproxeq 0$: Regge behaviour 
of $g_1(x,Q^2)$ is unstable against the QCD evolution. Let us mention for
comparison that in the spin-averaged case, $xF_1^s$ has the small $x$ 
behaviour as that
in Eq.\ref{asympt_g1} (in the Regge theory $F_1^s$ is controlled by the 
exchange of the pomeron with intercept $\sim$1.08) while $F_1^{ns}$ remains
stable under the QCD evolution ($F_1^{ns}$ is controlled by the exchange
of the $A_2$ trajectory of intercept $\sim$0.5).  

\begin{figure}[htb]
\begin{center}
\epsfig{figure=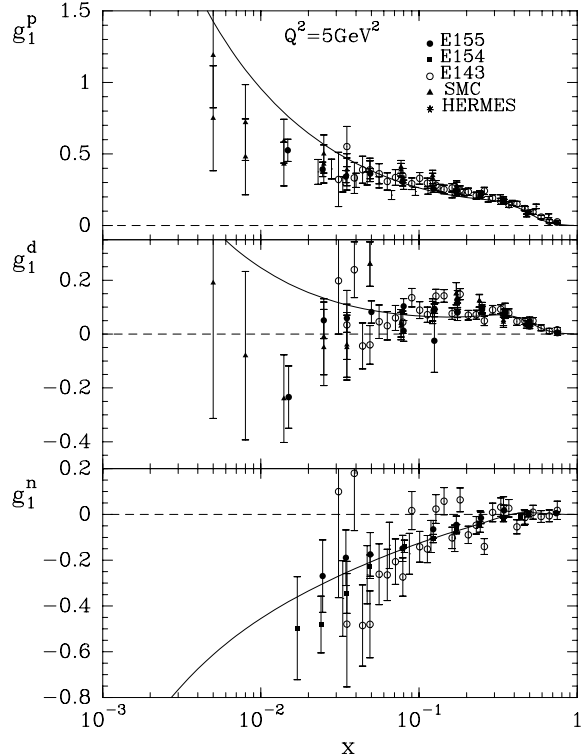,height=10cm}
\end{center}
\caption{\label{fig_soffer} \footnotesize{Spin dependent structure functions 
of proton, deuteron
and neutron from a global NLO QCD analysis in a statistical picture of 
the nucleon at $Q^2$ = 5 GeV$^2$ (curves).
The curves maintain their behaviour at least down to 
$x\sim$10$^{-5}$. Figure taken from \cite{qcd_soffer}.}}
\end{figure}

Several analyses of the $Q^2$ dependence of $g_1$ have been performed on 
the world data \cite{qcd_old,roland,grsv,qcd_bb,sidorov,qcd_soffer}, 
in the framework of the NLO QCD. However the present data do not permit to 
determine the shapes of parton distributions with sufficient 
accuracy. This is true especially for the small $x$ behaviour of parton
densities where neither the measurements nor the calculations of
possible new dynamic effects exist. Thus extrapolations of the DGLAP fit results 
to the unmeasured low $x$ region give different $g_1$ behaviours 
in different analyses, e.g. $g_1^p$ at $x\lapproxeq$
0.001 is positive and increasing with decreasing $x$ in \cite{qcd_soffer}, Fig.
\ref{fig_soffer} and negative and decreasing in \cite{qcd_old,grsv}. 
It should be stressed that
the $g_1$ results  for $x$ values below these of the data do not influence
the results of the fit. Thus there is no reason to expect that
the partons at very low $x$ behave as those in the measured (larger $x$) region.
Nevertheless extrapolations of the QCD fit are
presently being used to get the $x\rightarrow$ 0 extrapolation of $g_1$
 \cite{qcd_old}, necessary to evaluate its first moments. They
strongly disagree with the Regge asymptotic form, cf. Fig. \ref{fig_g1extrapo}.

\subsection{Double logarithmic \dln ~corrections to \fsn}

The LO (and NLO) QCD evolution which sums the powers of ln$(Q^2/Q^2_0)$ 
is incomplete at low $x$. Powers of another large logarithm, ln$(1/x)$, have to
be summed up there. In the spin-independent case this is accomplished by the
BFKL evolution equation (see e.g. \cite{bfkl}) which gives the leading low $x$
behaviour of the structure function, e.g. $F_1^s\sim x^{-\lambda_{BFKL}}$
where $\lambda_{BFKL}>$1.
   
It has recently been pointed out that the small $x$ behaviour of both singlet
and non-singlet
spin dependent structure function $g_1(x,Q^2)$ is controlled by the double
logarithmic terms, i.e. by those terms of the perturbative expansion which 
correspond to powers of
$\alpha_s {\rm ln}^2(1/x)$ at each order of the expansion, \cite{bartels}. 
The double logarithmic terms also
appear in the non-singlet spin averaged structure function $F_1^{ns}$
 \cite{MANA} but the leading small $x$ behaviour
of the $F_1^{ns}$ which they generate is
overriden by the (non-perturbative) contribution of the $A_2$ Regge
pole, \cite{jk_spin}. In case
of the $g_1$ its Regge behaviour is unstable against the resummation of the
${\rm ln}^2(1/x)$ terms which generate more singular $x$ dependence
than that implied by Eq.(\ref{regge_g1}) for $\alpha_{s,ns} \lapproxeq
0$, i.e. they generate the leading small $x$ behaviour of the $g_1$. 

The double logarithmic terms in the non-singlet part of the $g_1(x,Q^2)$
 are generated by ladder diagrams \cite{GORSHKOV,JKNS} as in Fig.
\ref{fig_ladder}. Contribution of non-ladder diagrams \cite{bartels} in
the non-singlet case is  non-leading in
the large $N_c$ limit ($N_c$ is a number of colours); it is numerically small
for $N_c$=3.
The contribution of non-ladder diagrams is however non-negligible for the
singlet spin dependent structure function; they are obtained from 
the ladder ones by adding to them soft bremsstrahlung gluons or soft quarks,
\cite{kz}. At low $x$, the singlet part, $g_1^s$ dominates $g_1^{ns}$.

The double logarithmic ${\rm ln}^2(1/x)$ effects go beyond the standard
LO (and NLO) QCD evolution of spin dependent parton densities.
They can be accomodated for in the QCD
evolution formalism based upon the renormalisation group equations, \cite{BLUM}. 
An alternative approach is based on unintegrated spin dependent parton 
distributions, $f_j(x^{\prime},k^2)$ ($j=u_{v},d_{v},\bar u,\bar d,\bar s,g$) where
$k^2$
is the transverse momentum squared of the parton $j$ and $x^{\prime}$
the
longitudinal momentum fraction of the parent nucleon carried by a parton
\cite{bbjk,kz,jk_app}. This formalism is very suitable for extrapolating 
$g_1$ to the region of low $Q^2$ at fixed $W^2$, \cite{bbjk}.    

The conventional (integrated) distributions $\Delta p_j(x,Q^2)$ (i.e.
$\Delta q_u=\Delta p_{u_v} + \Delta p_{\bar u}, ~
\Delta \bar q_u = \Delta p_{\bar u}$ etc. for quarks, antiquarks
and gluons) are
related in the
following way  to the unintegrated distributions $f_j(x^{\prime},k^2)$:
\begin{equation}
 \Delta p_j(x,Q^2)=\Delta p_j^{0}(x)+
 \int_{k_0^2}^{W^2}{dk^2\over k^2}f_j(x^{\prime}=
x(1+{k^2\over Q^2}),k^2)
\label{dpi}
\end{equation}
Here $\Delta p_j^{0}(x)$ denote the
nonperturbative parts of the of the distributions, corresponding to
$k^2 < k^2_0$ and the parameter $k_0^2$ is the infrared cut-off
($k_0^2 \sim $1 GeV$^2$). In \cite{bbjk,jk_app,kz} they were treated 
semiphenomenologically and were parametrised as follows:
\begin{equation}
\Delta p_j^0(x) = C_j (1-x)^{\eta_j}
\label{input}
\end{equation}   
The unintegrated distributions $f_j(x^{\prime},k^2)$ are the solutions of the
integral equations \cite{bbjk,jk_app,kz} which embody both the LO Altarelli-Parisi 
 evolution  and the double
ln$^2(1/x^{\prime})$ resummation at small $x^{\prime}$.
These equations combined with 
Eq.(\ref{dpi}) 
and with a standard relation of $g_1$ to the 
polarised quark and antiquark distributions $\Delta q_i$
and $\Delta \bar q_i$ corresponding to the quark (antiquark)
flavour $i$:
\begin{equation}
g_1(x,Q^2) = {1\over 2}\sum_{i=u,d,s} e_i^2\left[\Delta q_i(x,Q^2) +
\Delta \bar q_i(x,Q^2)
\right].
\label{gp1}
\end{equation}
(assuming $\Delta \bar q_{u} =\Delta \bar q_{d}$ and 
number of flavours equal 3) 
lead to approximate $x^{-\lambda}$ behaviour of the $g_1^{}$ in
the $x \rightarrow 0$ limit, with $\lambda \sim 0.4$ and
$\lambda \sim 0.8$ for the nonsinglet and singlet parts
respectively which is more singular at low $x$ than
that   generated by the (nonperturbative)  Regge pole exchanges.
%
\begin{figure}[htb]
\begin{center}
\epsfig{figure=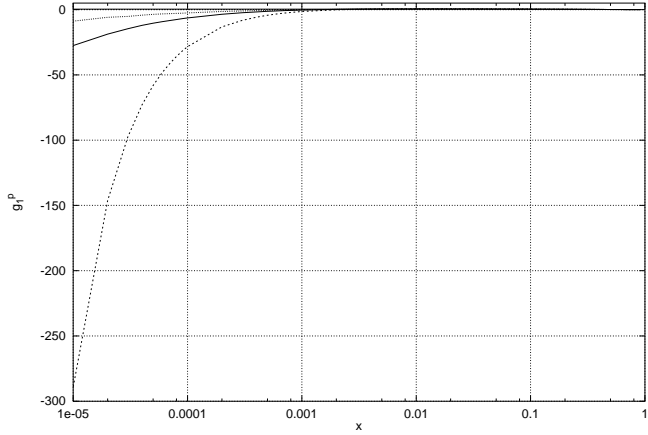,height=6cm}
\end{center}
\caption{\label{fig_kz} \footnotesize{
$g_1^p(x,Q^2)$ at $Q^2$=10 GeV$^2$. A thick solid line
corresponds to full calculations, a dashed one -- only the ladder ln$^2(1/x)$
resummation with LO Altarelli-Parisi evolution, a dotted one - pure LO
Altarelli-Parisi evolution and a thin solid line - the nonperturbative
input, $g_1^{(0)}$, related to $\Delta p_j^{0}(x)$ in Eq.\ref{dpi}. Figure
taken from \cite{kz}.
}}

\end{figure}

\begin{figure}[htb]
\begin{center}
\epsfig{figure=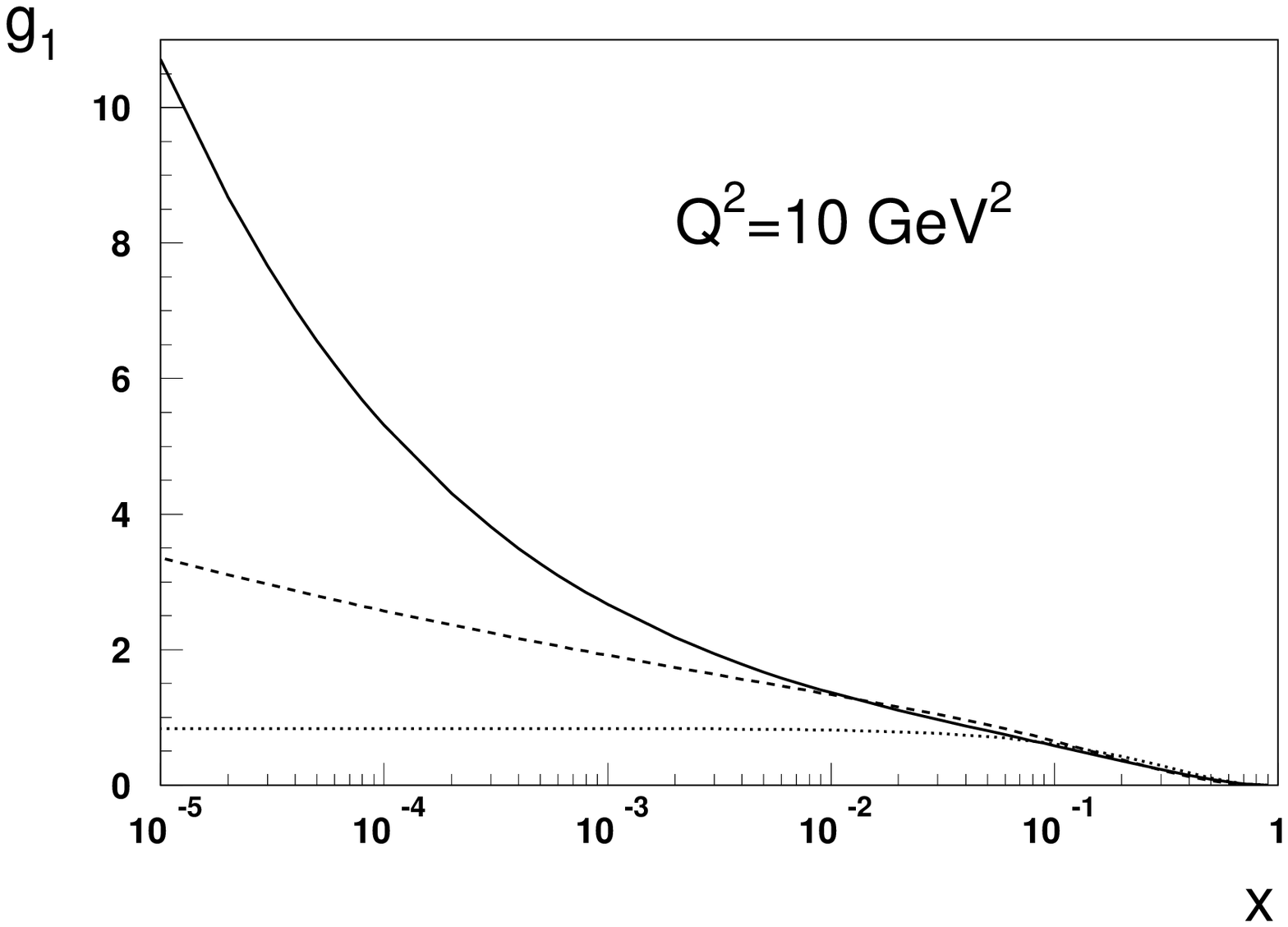,height=9cm}
\end{center}
\vskip-1.7cm
\caption{\label{fig_bbjk} \footnotesize{
Non-singlet part of the proton spin structure function $g_1(x,Q^2)$
at $Q^2=$10 GeV$^2$. Continuous line
corresponds to full calculations,
broken line is a pure leading order Altarelli--Parisi prediction,
and a dotted one marks the nonperturbative input, $g_1^{(0)}$, 
assumed as $g_1^{(0)}(x)=2g_A(1-x)^3/3$ and satisfying the Bjorken sum rule,
$\Gamma_1=g_A/6$. Figure taken from \cite{bbjk}.
}}
\end{figure}

Results of a complete, unified formalism incorporating
the LO Altarelli--Parisi evolution and the ln$^2(1/x)$ resummation at low $x$
for $g_1^p$ are shown in Figs \ref{fig_kz} and \ref{fig_bbjk},
separately for the total \cite{kz} and nonsinglet 
\cite{bbjk} parts of the spin dependent structure function. Resummation
of ln$^2(1/x)$ terms gives $g_1$ steeper than that generated by the
LO evolution alone and this effect is in $g_1^{ns}$ visible already for
$x\lapproxeq$ 10$^{-2}$.

The double ln$^2(1/x)$ effects are not important
in the $W^2$ range of the fixed target experiments. 
However since $x(1+k^2/Q^2) \rightarrow k^2/W^2$ for $Q^2 \rightarrow 0$
in the integrand in Eq. (\ref{dpi}) and since $k^2 >k_0^2$ there,
the $g_1^{}(x,Q^2)$ defined by Eqs (\ref{gp1}) and (\ref{dpi})
can be smoothly extrapolated to
the low $Q^2$ region, including $Q^2=0$. In that limit, the
$g_1$ should be a finite function of $W^2$, free from any
kinematical singularities or zeros. The extrapolation,
valid for fixed and large $W^2$, can thus be done  
provided that nonperturbative parts of the parton distributions
$\Delta p_j^{0}(x)$ are free from
kinematical singularities at $x=0$, as in the parametrisations
defined by Eq. (\ref{input}). If however $\Delta p_j^{0}(x)$  contain kinematical
singularities at $x$=0 then one may replace it with $\Delta p_j^{0}(\bar x)$
where $\bar x = x(1+k^2_0/Q^2)$ and leave the remaining parts of the calculations
unchanged.

The formalism including
the ln$^2(1/x)$ resummation  and the LO
Altarelli-Parisi evolution,
\cite{kz}, was used to calculate $g_1$ at $x$ and $Q^2$ values of the SMC 
measurement and a reasonable description of the data on $g_1^{p,d}(x,Q^2)$
extending down to $x\sim$0.0001 at $Q^2\sim$0.02 GeV$^2$
was obtained, cf. Fig.1 in \cite{bkk}. 
Of course the (extrapolated) partonic contribution 
may not be the only one at low $Q^2$; the VMD part may play a non-negligible role as
well, cf. Section 5.

\subsection{Low $x$ contributions to $g_1$ moments}
Fundamental tools in investigating the
properties of the spin interactions are the sum rules, expected to be satisfied by 
the spin structure functions.
These sum rules involve first moments of $g_1$, i.e. integrations of  $g_1$ over the 
whole range of $x$ values, from 0 to 1. This means that experimentally unmeasured
regions, [0,$x_{min}$) and ($x_{max}$,1] must also be included in the integrations.
The latter is not critical, see e.g.\cite{roland_exp}, but contribution from
the former may significantly influence the moments. The value of $x_{min}$ 
depends on the value of the maximal lepton energy loss, $\nu_{max}$, accessed
in an experiment at a given $Q_0^2$. For the CERN experiments, with muon beam 
energies about 200 GeV and at $Q_0^2$=1 GeV$^2$ it is about 180 GeV which
corresponds to $x_{min}\approx$ 0.003. Contribution to the 
$g_1$ moments from the unmeasured region, 0$\leq x <$ 0.003, has thus to be done
phenomenologically.  
  
Unified system of equations including
the double ln$^2(1/x)$ resummation effects and the complete leading-order
Altarelli-Parisi evolution,
\cite{kz}, was used to extrapolate the spin dependent parton distributions 
and the polarised nucleon structure functions down to  $ x \sim 10^{-5}$, 
\cite{beata}. Calculated moments of the proton stucture function for 
2 $< Q^2 <$ 15 GeV$^2$, i.e. where the low $x$ measurements exist, agreed well with
the latter and the estimated contribution of the integral over $g_1(x,Q^2)$ in the
interval 10$^{-5} < x < 10^{-3}$ was about 2\% of the total $g_1^p$ moment 
in the above interval of $Q^2$. In the same limits of $Q^2$, moments of 
$g_1^n$ were found to lie below the experimental data and the calculated low $x$ 
contribution was 8\% of the total neutron moment. All these contributions 
increase with increasing $Q^2$. It was also estimated that 
the above low $x$ region contributes only 
in about 1\% and 2\% to the Bjorken and Ellis--Jaffe sum rules respectively.   

Within the same formalism and at $Q^2$=10 GeV$^2$, a contribution of 0.0080 
from the unmeasured region, 0$\leq x <$ 0.003, to the Bjorken integral was obtained 
while the contribution resulting from
the pure LO Altarelli-Parisi evolution
was 0.0057. These numbers have to be compared with 0.004 obtained when $g_1$=const
behaviour, consistent with Regge prediction was assumed and fitted to the
lowest $x$ data points for proton and deuteron targets (see \cite{bbjk} and
references therein). 

Extrapolation to the unmeasured region (0$\leq x <$ 0.003) of the NLO DGLAP fits 
to the world data 
results in about 10\% contribution of that low $x$ region to the $g_1^p$ 
moment, \cite{qcd_old}. 
The NLO DGLAP fit to the SMC data gave a contribution of 0.010 to the Bjorken 
integral at $Q^2$=10 GeV$^2$, 
i.e. about 6\% of that integral, \cite{qcd_old}.
These numbers rely on the validity of the assumption that 
the parton distributions behave as $x^\delta$ as $x\rightarrow$0.
 
\section{Nonperturbative effects in $g_1$}
Data on polarized nucleon structure function $g_1(x,Q^2)$
extend to the region of low values of 
$Q^2$, \cite{nne143,smc98,ssmc98,uta}.
This region is of particular interest since
nonperturbative mechanisms dominate the particle dynamics there and 
a transition from soft- to hard physics may be studied.
In the fixed target experiments the low values
of $Q^2$ are reached simultaneously with the low values of the Bjorken variable, 
$x$, cf. Figs \ref{fig_t15} and \ref{fig_compass}, and therefore predictions for
spin structure functions in both the low $x$
and low $Q^2$ region are needed. Partonic contribution to $g_1$ which controls
the structure function in the deep inelastic domain
has thus to be suitably extended to the
 low $Q^2$ region and complemented by a non-perturbative component. 

The low $Q^2$ spin-averaged electroproduction is very successfuly described by
the Generalised Vector Meson Dominance (GVMD) model, 
see e.g. \cite{breitweg,el89,el92}.
Therefore me\-thods based on GVMD should also be used to describe the behaviour of
the $g_1$ in the low $x$, low $Q^2$ region. Two attempts using such methods
have recently been made. In the first one, \cite{bkk} 
the following representation of $g_1$ was assumed:
\eq
g_1(x,Q^2)=g_{1}^{VMD}(x,Q^2) + g_{1}^{part}(x,Q^2).
\label{g1tot}
\eqx
The partonic contribution, $g_{1}^{part}$ which at low $x$ is controlled
by the logarithmic ln$^2(1/x)$ terms, was parametrised as discussed in Section 4.2. 

The VMD contribution, $g_1^{VMD}(x,Q^2)$, was represented as:
\noindent
\eq
g_{1}^{VMD}(x,Q^2)  = {M\nu\over 4 \pi} \sum_{V=\rho,\omega,\phi}
{M_V^4 \Delta \sigma_{V}(W^2) \over \gamma_V^2
(Q^2+M_V^2)^2}
\eqx
where $M_V$ is the mass of the vector meson $V$,
$\gamma_V^2$ are determined from the leptonic widths of the vector mesons
 and $\nu=Q^2/2Mx$.
The unknown cross sections $\Delta \sigma _{V}(W^2)$ are combinations
of the total cross sections for the scattering of polarised vector mesons 
and nucleons. It was assumed that they are proportional
(with a proportionality coeffcient $C$)
to the appropriate combinations of the nonperturbative contributions
$\Delta p_j^0(x)$ to the polarised quark and antiquark distributions:
$$
{M\nu\over 4 \pi} \sum_{V=\rho,\omega}{M_V^4 \Delta \sigma_{V}
\over \gamma_V^2 (Q^2 + M_V^2)^2} =
$$
\begin{equation}
C \left  [ {4\over 9} \left(\Delta u^0_{val}(x) + 2\Delta \bar u^{0}(x)\right)
+{1\over 9} \left(\Delta d^{0}_{val}(x) + 2\Delta \bar d^0(x)\right)\right]
{M_{\rho}^4\over (Q^2+M_{\rho}^2)^2},
\label{nsvmd}
\end{equation}
\begin{equation}
{M\nu\over 4 \pi} {M_{\phi}^4 \Delta \sigma_{\phi p} \over \gamma_{\phi}^2
 (Q^2 + M_{\phi}^2)^2} =
C{2\over 9}\Delta \bar s^0(x){M_{\phi}^4 \over (Q^2 +M_{\phi}^2)^2},
\label{svmd}
\end{equation}   
where $\Delta u^0(x)=\Delta p_u^{0}(x)$, etc.
The $\Delta p^0_j(x)$, Eq.(\ref{input}), behave as $x^0$ for $x\rightarrow$0.
As a result the  cross sections $\Delta \sigma_{V}$ behave as
$1/W^2$ at large $W^2$ which corresponds to zero intercepts of
the appropriate Regge trajectories.
\begin{figure}[ht]
\epsfig{figure=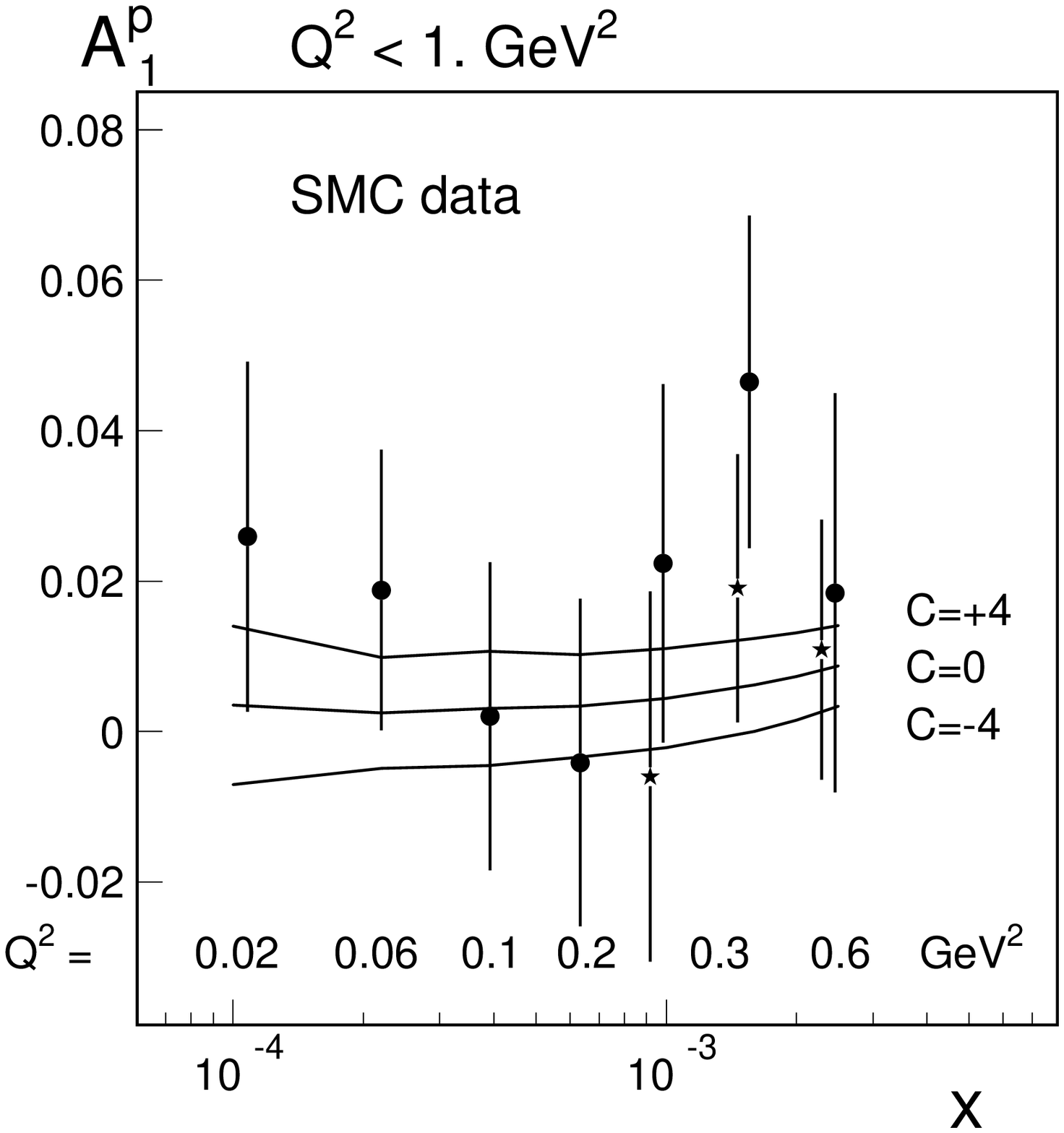,height=7cm}
\hspace*{0.5cm}
\epsfig{figure=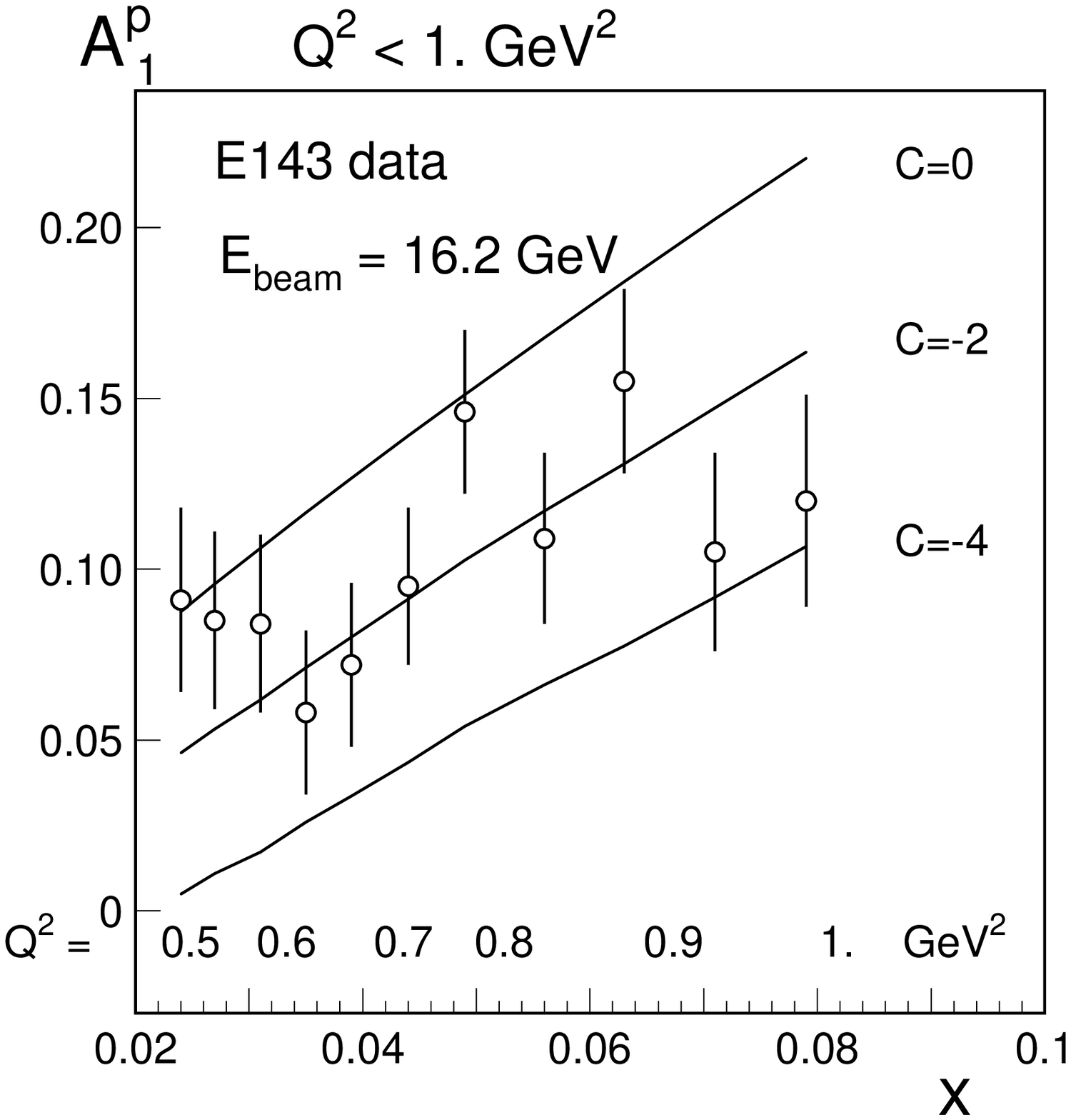,height=7cm}
\caption{\label{fig_a1p} \footnotesize{
Spin asymmetry $A_1$ for the proton as a function
of $x$ at the measured $Q^2$ values (marked above the $x$ axis),
obtained by the SMC (stars \protect\cite{smc98} and dots \protect\cite{ssmc98})
and by SLAC E143 \protect\cite{nne143} (at 16.2 GeV incident energy only).
Errors are statistical. Curves are predictions of the model for
different values of $C$. Figure comes from  \protect\cite{bkk}.}
}
\end{figure}

Results of calculations for $Q^2<$1 GeV$^2$ are shown in Fig.\ref{fig_a1p} 
for different values of $C$. 
The statistical accuracy of the SMC data is too poor to constraint
the value of the coefficient $C$. The SLAC E143 data apparently prefer a small
negative value of $C$.  Similar analysis of the neutron and deuteron spin
structure functions was inconclusive.

In the other attempt to describe the \fsn ~in the low $x$, low $Q^2$ region, 
\cite{bkz}, the GVMD model was used together 
with the Drell-Hearn-Gerasimov-Hosoda-Yamamoto (DHGHY) sum rule, \cite{DH,GER,HY}. 
In the GVMD, the \fsn ~has the following representation, valid for fixed 
$W^2\gg Q^2$, i.e. small values of $x$, $x=Q^2/(Q^2+W^2-M^2)$:
{\footnotesize
\eq
g_1(x,Q^2)=g_1^{L}(x,Q^2)+g_1^{H}(x,Q^2)=\frac{M\nu}{4\pi}\,\sum_V\,
\frac{M^4_V \Delta \sigma_V(W^2)}{\gamma_V^2(Q^2+M^2_V)^2}+
g_1^{AS}({\bar x},Q^{2}+Q_0^2).
\label{gvmd}
\eqx
}
\noindent
The first term
sums up contributions from light vector mesons, $M_V < Q_0$ where  $Q_0^2
\sim$ 1 GeV$^2$ \cite{el89}. The unknown $\Delta \sigma_V$ are
expressed through the combinations of nonperturbative parton distributions,
$\Delta p_j^{0}(x)$, evaluated at fixed $Q_0^2$, similar to the previous
case.

The second term in (\ref{gvmd}), $g_1^{H}(x,Q^2)$, which represents
the contribution of heavy ($M_V > Q_0$) vector mesons to $g_1(x,Q^2)$
can also be treated as an extrapolation of the QCD improved parton model
structure function, $g_1^{AS}(x,Q^{2})$, to arbitrary values of $Q^2$:
$g_1^H(x,Q^2)=g_1^{AS}(\bar x,Q^2+Q_0^2)$, cf. \cite{el92}. Here
the scaling variable $x$ is replaced by ${\bar x}=(Q^2+Q_0^2)/(Q^2+Q_0^2+W^2-M^2)$. 
It follows  that $g_1^{H}(x,Q^2)\rightarrow g_1^{AS}(x,Q^{2})$ as $Q^2$ is large.
We thus get:
{\footnotesize
\eqn
g_1(x,Q^2)&=&
%
C\left[ \frac{4}{9}(\Delta u_{val}^{0}(x)+2\Delta \bar u^{0}(x))
+\frac{1}{9}(\Delta d_{val}^{0}(x)+2\Delta \bar d^{0}(x))\right]
\frac{M^4_{\rho}}{(Q^2+M^2_{\rho})^2}\nonumber\\
&+&C\left[ \frac{1}{9}(2\Delta \bar s^{0}(x))\right]\frac{M^4_{\phi}}
{(Q^2+M^2_{\phi})^2}\nonumber \\
&+&g_1^{AS}({\bar x},Q^{2}+Q_0^2).\label{gvmdu}
\eqnx
}
\noindent
The only free parameter in (\ref{gvmdu}) is the constant $C$. Its value
may be fixed in the photoproduction limit where the first moment of
$g_1(x,Q^2)$ is related to the anomalous magnetic moment of the
nucleon via the DHGHY sum rule, cf.\ \cite{ioffe,ioffe2}:
\eq
I(0)=I_{res}(0)
+ M\,\int_{\nu_t(0)}^{\infty}\,\frac{d\nu}{\nu^2}\,g_1\left(x(\nu),0\right)
=-\kappa^2_{p(n)}/4,
\label{i0}
\eqx
where the DHGHY moment before taking the $Q^2$=0 limit has been split into 
two parts, one corresponding
to $W < W_t\sim$ 2 GeV (baryonic resonances) and the other to $W > W_t$:
\eq
I(Q^2)=I_{res}(Q^2)
+ M\,\int_{\nu_t(Q^2)}^{\infty}\,\frac{d\nu}{\nu^2}\,g_1\left(x(\nu),Q^2\right).
\label{iq2}
\eqx
Here $\nu_t(Q^2)=(W_t^2+Q^2-M^2)/2M$. Substituting $g_1\left(x(\nu),0\right)$
in Eq. (\ref{i0}) by Eq. (\ref{gvmdu}) at $Q^2=0$
we may obtain the value of $C$ from (\ref{i0}) if $I_{res}(0)$, the contribution
from resonances, is known e.g. from measurements.

%
%
%
\begin{figure}[ht]
\begin{center}
\vspace*{-1cm}
\epsfig{width=12cm, file=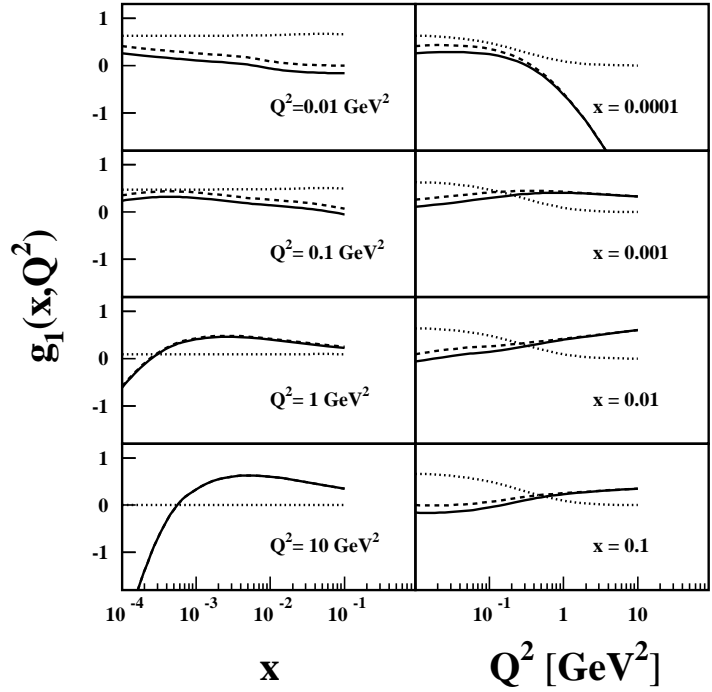}
\vspace*{-1.5cm}
\end{center}
\caption{\label{dhg_g1}\footnotesize{Values of $xg_1$ for the proton 
as a function of $x$ and $Q^2$, Eq.(\ref{gvmdu}). The asymptotic contribution, 
$g_1^{AS}$,
is marked with broken lines, the VMD part, $g_1^{\rm L}$, with dotted lines
and the continuous curves mark their sum, according to (\ref{gvmdu}). 
Figure comes from \cite{bkz}.
}
}
\end{figure}
To obtain the value of $C$ from Eq.
(\ref{i0}), $I_{res}(0)$ was evaluated using the preliminary
 data taken at ELSA/MAMI by the GDH Collaboration \cite{elsa}
at the photoproduction, for $W_t$=1.8 GeV. The $g_1^{AS}$
was parametrized using GRSV fit \cite{grsv} for the ``standard
scenario'' at the NLO accuracy. 
$Q_0^2$ = 1.2 GeV$^2$ was assumed as in the analysis of $F_2$, \cite{el89}.
As a result the constant $C$ was found to be --0.24 or 
--0.30, for the $\Delta p_j^{0}(x)$ in Eq.(\ref{gvmdu}) parametrised 
at $Q^2=Q_0^2$ as
Eq. (\ref{input}) or as \cite{grsv}, respectively. 

The nonperturbative, Vector Meson Dominance
contribution was obtained negative in both attempts, \cite{bkk,bkz}
as well as  from earlier phenomenological analyses
of the sum rules \cite{ioffe2,burkert}.

\begin{figure}[ht]
\vspace*{1cm}
\hspace*{0.5cm}
\epsfig{width=7cm, file=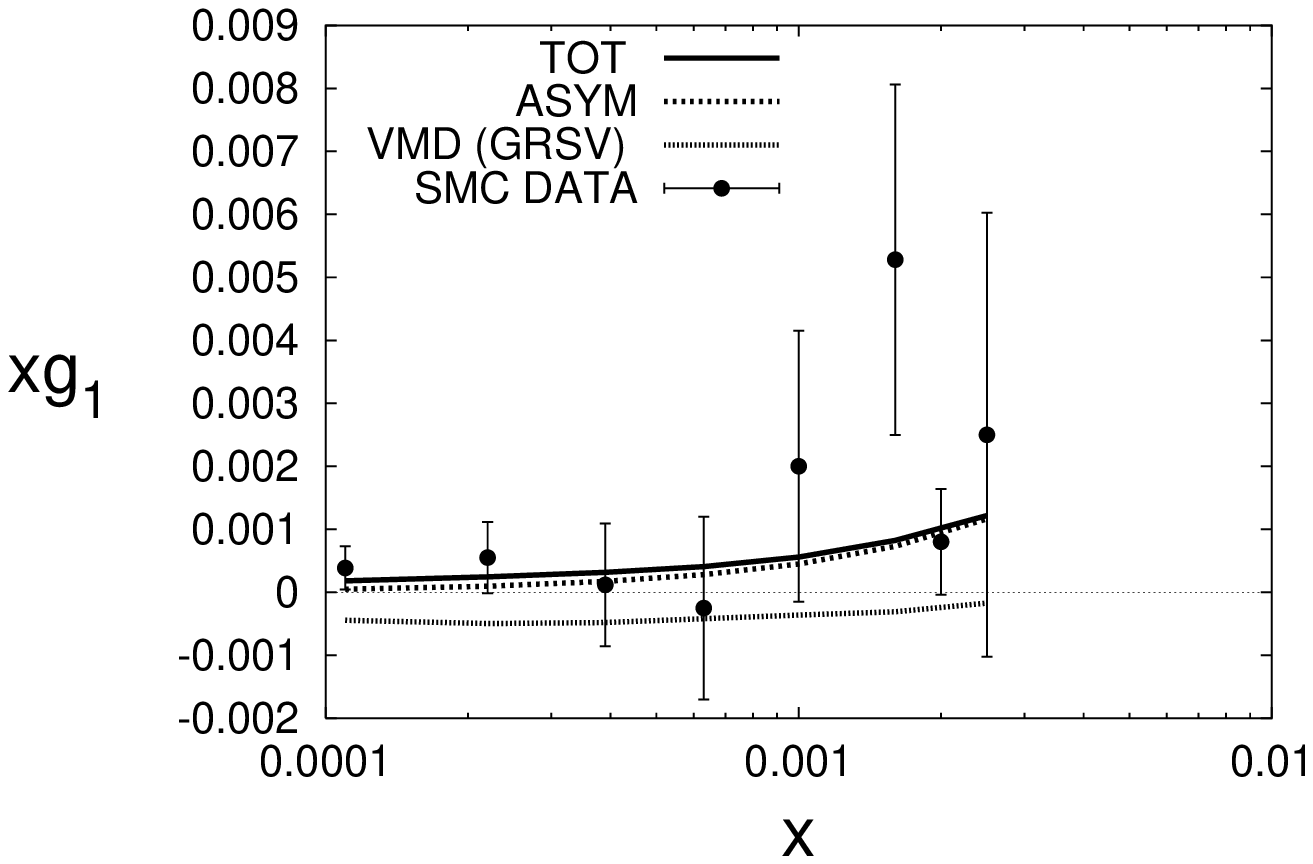}
\vspace*{2cm}
\hspace*{0.5cm}
\epsfig{width=7cm, file=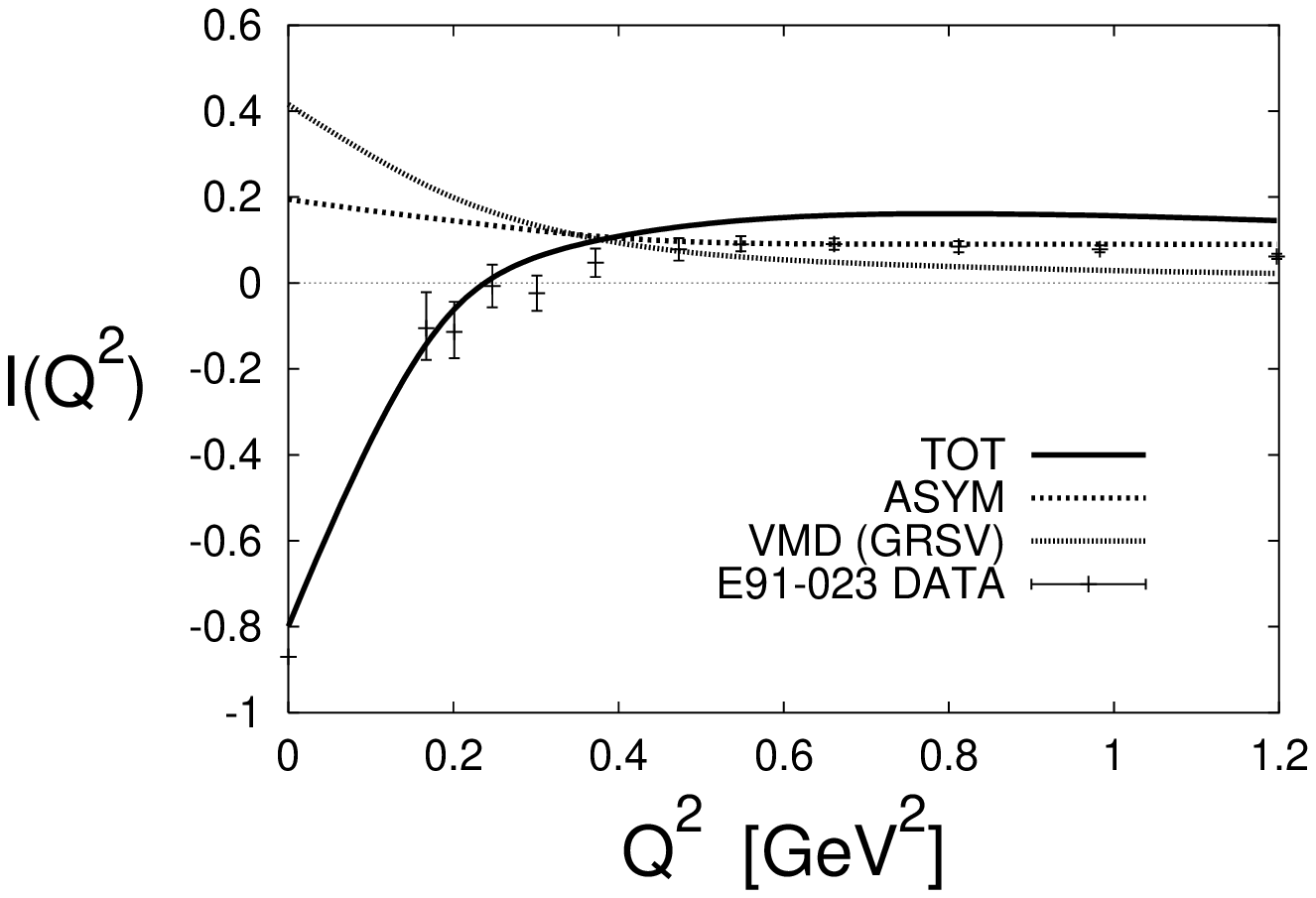}
\vspace*{-2cm}
\caption{\label{fig1}\footnotesize{a) Values of $xg_1$ for the proton as a function of $x$ at the
measured values of $Q^2$ in the non-resonant region, $x<x_t=Q^2/2M\nu_t(Q^2)$.
Both the VMD input and $g_1^{AS}$ have been evaluated using the GRSV
fit for standard scenario at the NLO accuracy \cite{grsv}.
Contributions of the VMD and of the $xg_1^{AS}$
are shown separately. Points are the SMC measurements at
$Q^2 <$ 1 GeV$^2$, \cite{ssmc98}; errors are total. The
curves have been calculated at the measured $x$ and $Q^2$ values.
b) The DHGHY moment $I(Q^2)$ for the proton. Details as in Fig.\ref{fig1}a.
Points mark the contribution of resonances as measured by the JLAB E91-023,
\cite{e91-023} at $W < W_t(Q^2)$. Figures come from \cite{bkz}.}
}
\end{figure}

\begin{figure}[ht]
\begin{center}
\epsfig{width=8cm, height=5.3cm, file=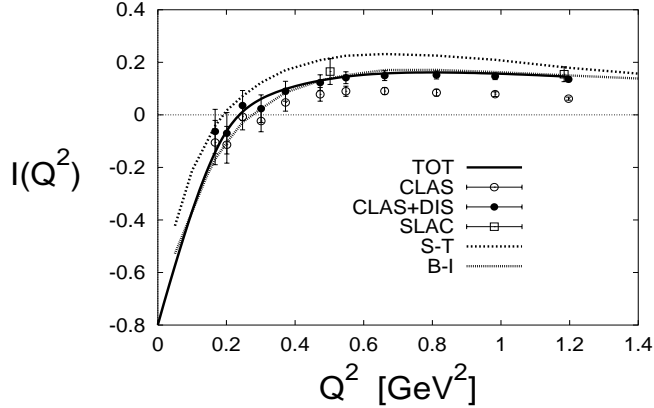}\\
\end{center}
\caption{\label{fig3}\footnotesize {The DHGHY moment $I(Q^2)$ for the proton with the VMD part 
parametrized using the GRSV fit \cite{grsv}.
Shown are also calculations of \cite{burkert}
(``B--I'') and \cite{soffer} (``S--T''). Points marked ``CLAS'' are from
the JLAB E91-023 experiment
\cite{e91-023}: the open circles refer to the resonance region, $W < W_t(Q^2)$
and the full circles contain a correction for the DIS contribution.
Errors are total. Figure comes from \cite{bkz}.}
}
\end{figure}

The $g_1$ obtained from the above formalism is shown in Fig. \ref{dhg_g1}.
It reproduces well a general trend in the data, cf. 
Fig.\ref{fig1}a; however experimental errors are too large for
a more detailed analysis. To compute the DHGHY moment, Eq.(\ref{iq2}),
for the proton, preliminary results of the JLAB E91-023
experiment \cite{e91-023} for 0.15$\lapproxeq Q^2 \lapproxeq $1.2 GeV$^2$
and $W < W_t=W_t(Q^2)$~\cite{fatemi} were used. Results, Fig.\ref{fig1}b, show that
partons contribute significantly even at $Q^2\rightarrow$ 0 where the
main part of the $I(Q^2)$ comes from resonances. 
The DHGHY moment is shown in Fig. \ref{fig3} together with the results
of calculations of Refs \cite{burkert,soffer} as well as with
the E91-023 measurements in the resonance region used as an input to the
$I(Q^2)$ calculations. The E91-023 data
corrected by their authors for the deep inelastic contribution are also presented.
Results of calculations are slightly
larger than the DIS-corrected data and the results of \cite{burkert} but
clearly lower than the results of \cite{soffer} which overshoot the data.\\
\noindent
\vskip-1cm
%
%

\section{Outlook}

The longitudinal spin dependent structure function, \fsn, is presently the only
observable which permits an insight into the spin dependent low $x$ physics.
Contrary to spin-independent structure functions, it is sensitive to double
logarithmic, ln$^2(1/x)$ corrections, generating its leading small $x$ behaviour.
However its knowledge is limited by the statistical accuracy 
and by the kinematics of the fixed-target experiments. In the latter, the
low values of $x$ are reached simultaneously with the low values of the four 
momentum transfer, $Q^2$. While the low $Q^2$ domain may be of great interest due
to a transition from soft to hard physics, it also challenges theoretical
predictions based on partonic ideas which have to be suitably extended 
to the nonperturbative region.

Until now, experimental data on the \fsn ~at low $x$ came mainly from the SMC at CERN.
They do not permit to constrain the low $x$ parton distributions, nor
to test the Regge model but they seem to leave room for contributions
other than (low $Q^2$ extrapolated) partonic mechanisms. They also permitted
first quantitative studies of nonperturbative mechanisms; results consistently
point towards large and negative contribution of the latter. 

New low $x$ data on \fsn
~will soon be available from COMPASS. Their statistics will be by far 
larger so that statistical errors should no longer be do\-mi\-nating. Also
the experimental acceptance at low $x$ will be much wider in the 
nonperturbative domain and thus tests of the Regge behaviour of $g_1$ 
will be possible. A crucial extension of the kinematic domain of the
(deep) inelastic spin electroproduction will take place with the advent of
the polarised Electron-Ion Collider, EIC, at BNL \cite{eic,hera_study_lowq}. With its 
centre-of-mass energy only about 2 times lower than that at HERA, this machine will 
open a field of pertubative low $x$ spin physics where also other,  
semi-inclusive and exclusive observables, will be accessible for testing the high 
parton density mechanisms.       

 
\section*{Acknowledgments} 
Many subjects in low $x$ physics were for the first time addressed
and developed by Jan Kwieci\'nski. One of these is the  non-perturbative 
aspect of the low $x$ electroproduction . His involvement in the phenomenology
of this exciting and still not understood branch of high energy physics
was stimulated by the EMC measurements, followed by results of NMC, E665, 
SMC and HERA. I am greatly indebted to Jan for many years of most
enjoyable scientific collaboration.  

This research has been supported in part by the Polish Committee for Scientific
Research with grant number 2P03B 05119.
  
%


\begin{thebibliography}{9999}
\bibitem{martin}A. Martin, in ``Spin in physics'', X S\'eminaire Rhodanien de 
Physique, Villa Gualino, March 2002, edited by M. Anselmino, F. Mila and
J. Soffer, Frontier Group.
\bibitem{future}J. Soffer, hep-ph/0212011, closing talk at the 15th International
Spin Physics Symposium (SPIN2002), Long Island, NY, September 2002 (to appear
in the Proceedings).
\bibitem{heraf2} ZEUS Collaboration, S. Chekanov et al., Eur. Phys. J. {\bf C21}
(2001) 443; H1 Collaboration, C. Adloff et al., Phys. Lett. {\bf B520} (2001) 183.
\bibitem{lowx_coll}Small $x$ Collaboration, B. Andersson et al.,
Eur. Phys. J. {\bf C25} (2002) 77.
\bibitem{breitweg} ZEUS Collaboration, J. Breitweg et al., Phys. Lett. 
{\bf B487} (2000) 53 and references therein. 
\bibitem{el89} B. Bade\l ek and J. Kwieci\'nski, Z. Phys. {\bf C43} (1989) 251.
\bibitem{el92} B. Bade\l ek and J. Kwieci\'nski, Phys. Lett. {\bf B295} (1992) 
263.
\bibitem{smc98} SMC, B. Adeva et al., Phys. Rev. {\bf D58} (1998) 112001.
\bibitem{ssmc98} SMC, B. Adeva et al., Phys. Rev. {\bf D60} (1999) 072004
and erratum: {\it ibid.} {\bf D62} (2000) 079902.
\bibitem{claude}COMPASS Collaboration, C. Marchand, talk at the XIth 
International Workshop on
Deep Inelastic Scattering (DIS2003), April 2003, St Petersburg, Russia;
to appear in the Proceedings.
\bibitem{roland_exp} R. Windmolders, in ``Spin in physics'', X S\'eminaire 
Rhodanien de Physique, Villa Gualino, March 2002, edited by M. Anselmino, 
F. Mila and J. Soffer, Frontier Group.

\bibitem{uta} U. St\"osslein, Proceedings of the Xth 
International Workshop on Deep Inelastic Scattering (DIS2002),
May 2002, Cracow, Poland, Acta Phys. Pol. {\bf B33} (2002) 2813.
 
\bibitem{hei} R.L. Heimann, Nucl. Phys. {\bf B64} (1973) 429;
J. Ellis and M. Karliner, Phys. Lett. {\bf B213} (1988) 73.
\bibitem{nicola} N. Bianchi and E. Thomas, Phys. Lett. {\bf B450} (1999) 439.
\bibitem{clo_rob} F.E. Close and R.G. Roberts, Phys. Lett. {\bf B336} (1994) 257.
\bibitem{bass_land} S.D. Bass and P.V. Landshoff, Phys. Lett. {\bf B336} (1994) 537.
\bibitem{misha} J. Kuti in ``The Spin Structure of the Nucleon'',
World Scientific, Singapore, 1996; M.G. Ryskin, private communication (Durham, 1998).
\bibitem{qcd_old} SMC, B. Adeva et al., Phys. Rev. {\bf D58} (1998) 112002.
\bibitem{hera_study} A. De Roeck et al., Eur. Phys. J. {\bf C6} (1999) 121.
\bibitem{bbjk} B. Bade\l ek and J. Kwieci\'nski, Phys. Lett. {\bf B418} (1998) 
229.
\bibitem{roland} For a review of early QCD analyses of $g_1$ see e.g.
R. Windmolders, Nucl. Phys. B (Proc. Suppl.) {\bf 79} (1999) 51.
\bibitem{grsv} M. Gl\"uck, E. Reya, M. Stratmann and W. Vogelsang, Phys. Rev.
{\bf D63} (2001) 094005.
\bibitem{qcd_bb} J. Bl\"umlein and H. B\"ottcher, Nucl. Phys. {\bf B636} (2002) 225. 
\bibitem{sidorov} E. Leader, A.V. Sidorov and D.B. Stamenov, Eur. Phys. J. 
{\bf C23} (2002) 479. 
\bibitem{qcd_soffer} C. Bourrely, J. Soffer and F. Buccella, Eur. Phys. J. {\bf C23}
(2002) 487.
\bibitem{bfkl}E.A. Kuraev, L.N. Lipatov and V. Fadin, Zh. Eksp. Teor. Fiz.
{\bf 72} (1977) 373 [Sov. Phys. JETP, {\bf 45} (1977) 199]; Ya. Ya. Balitzkii and 
L.N. Lipatov, Yad. Fiz. {\bf 28} (1978) 1597 [Sov. J. Nuc. Phys. {\bf 28} (1978) 822;
L.N. Lipatov, in ``Perturbative QCD'', edited by A.H. Mueller, World Scientific, 1989.
\bibitem{bartels}J. Bartels, B.I. Ermolaev and M.G. Ryskin, Z. Phys. 
{\bf C70} (1996) 273; {\it ibid.} {\bf C72} (1996) 627.
\bibitem{MANA} B.I. Ermolaev, S.I. Manayenkov and M.G. Ryskin,
Z. Phys. {\bf C69} (1996) 259; S.I. Manayenkov, Z. Phys. {\bf C75} (1997) 685.
\bibitem{jk_spin}J. Kwieci\'nski, Acta Phys. Pol. {\bf B27} (1996) 893.
\bibitem{GORSHKOV}V.G. Gorshkov {\it et al.}, Yad. Fiz. {\bf 6} (1967)
129 (Sov. J. Nucl. Phys. {\bf 6} (1968) 95); L.N. Lipatov, Zh.
Eksp. Teor. Fiz. {\bf 54} (1968) 1520 (Sov. Phys. JETP {\bf 27}
(1968) 814. R. Kirschner and L.N. Lipatov, Nucl. Phys. {\bf
B213} (1983) 122; R. Kirschner, Z. Phys. {\bf C67} (1995) 459.
\bibitem{JKNS} J. Kwieci\'nski, Phys. Rev. {\bf D26} (1982)
3293.
\bibitem{kz} J. Kwieci\'nski and B. Ziaja,  Phys. Rev. {\bf D60} (1999) 054004.
\bibitem{jk_app}J. Kwieci\'nski, Acta Phys. Pol. {\bf B29} (1998) 1201.
\bibitem{BLUM}
 J. Bl\"umlein and A. Vogt, Acta Phys. Polon. {\bf B27} (1996) 1309;
 Phys.Lett. {\bf B386} (1996) 350; J. Bl\"umlein, S. Riemersma and A. Vogt
 Nucl. Phys. B (Proc. Suppl)  {\bf 51C} (1996) 30.
\bibitem{bkk} B. Bade\l ek, J. Kiryluk and J. Kwieci\'nski,  Phys. Rev. {\bf D61}
 (2000) 014009.
\bibitem{beata} B. Ziaja, Acta Phys. Pol., {\bf B32} (2001) 2863. 
\bibitem{nne143}SLAC, E143, K. Abe et al., Phys. Rev. {\bf D58}
 (1998) 112003.


\bibitem{bkz}B. Bade\l ek, J. Kwieci\'nski and B. Ziaja,  Eur. Phys. J. {\bf C26}
(2002) 45.
\bibitem{DH} S.D. Drell and A.C. Hearn, Phys. Rev. Lett. {\bf 16} (1966) 
908.
5A
\bibitem{GER} S.B. Gerasimov, Sov. J. Nucl. Phys. {\bf 2} (1966) 430.
\bibitem{HY} M. Hosoda and K. Yamamoto, Prog. Theor. Phys. Lett. {\bf 36} (1966) 
425.
\bibitem{ioffe} B.L. Ioffe, Surveys in High Energy Physics, {\bf 8} (1995) 107.
\bibitem{ioffe2} B.L. Ioffe, Phys. At. Nucl. {\bf 60} (1997) 1707.
\bibitem{elsa} GDH Collaboration, K. Helbing, Nucl. Phys. (Proc. Suppl.) 
{\bf 105} (2002) 113.
\bibitem{burkert} V. Burkert and B.L. Ioffe, Phys. Lett. {\bf B296} (1992) 223;
V. Burkert and B.L. Ioffe,  J. Exp. Th. Phys. {\bf 78} (1994) 619.
\bibitem{e91-023} JLAB E91-023, R. De Vita, talk at BARYONS 2002,
Jefferson Lab., March 3-8, 2002, 
http://www.jlab.org/baryons2002/program.html. 
\bibitem{fatemi} JLAB E91-023, R. Fatemi, private communication (June 2002).
\bibitem{soffer} J. Soffer and O. Teryaev, Phys. Rev. Lett. {\bf 70} (1993) 3373.

\bibitem{eic}See e.g. http://www.bnl.gov/eic.
\bibitem{hera_study_lowq} S.D. Bass and A. De Roeck, Eur. Phys. J. {\bf C18} (2001) 
531.
\end{thebibliography}
\end{document}